\documentclass[twocolumn,showpacs]{revtex4}
\usepackage{color}
\usepackage{amssymb}
\usepackage{amsmath}
\usepackage{mathrsfs,dsfont}
\usepackage{bm}

\usepackage{graphicx}
\usepackage{natbib}

\newcommand{\comment}[1]{} 

\newcommand{\dg}{\delta_{\rm g}}
\newcommand{\dl}{\delta_{\rm lin}}
\newcommand{\Pl}{P_{\rm lin} } 
\newcommand{\hMpc}{h^{-1}\,\rm Mpc}

\newcommand{\mnras}{Mon. Not. R. Astron. Soc.}

\newcommand{\aap}{Astron. Astrophys.}
\newcommand{\apjl}{Astrophys. J. Lett.}
\newcommand{\jcap}{J. Cosmo. Astropart. Phys.}
\newcommand{\pasj}{Proc. Astron. Soc. Japan}
\newcommand{\physrep}{Phys. Rep.}

\allowdisplaybreaks

\begin{document}

\title{Streaming velocities and the baryon-acoustic oscillation scale}

\author{Jonathan A. Blazek}
\email{blazek@berkeley.edu}
\author{Joseph E. McEwen}

\author{Christopher M. Hirata}

\affiliation{Center for Cosmology and AstroParticle Physics, The Ohio State University, 191 West Woodruff Avenue, Columbus, Ohio 43210, USA}

\date{February 3, 2016}

\begin{abstract}

At the epoch of decoupling, cosmic baryons had supersonic velocities relative to the dark matter that were coherent on large scales. These velocities subsequently slow the growth of small-scale structure and, via feedback processes, can influence the formation of larger galaxies. We examine the effect of streaming velocities on the galaxy correlation function, including all leading-order contributions for the first time. We find that the impact on the BAO peak is dramatically enhanced (by a factor of $\sim 5$) over the results of previous investigations, with the primary new effect due to advection: if a galaxy retains memory of the primordial streaming velocity, it does so at its Lagrangian, rather than Eulerian, position. Since correlations in the streaming velocity change rapidly at the BAO scale, this advection term can cause a significant shift in the observed BAO position. If streaming velocities impact tracer density at the 1\% level, compared to the linear bias, the recovered BAO scale is shifted by approximately 0.5\%. This new effect, which is required to preserve Galilean invariance, greatly increases the importance of including streaming velocities in the analysis of upcoming BAO measurements and opens a new window to the astrophysics of galaxy formation.

\end{abstract}

\maketitle

\section{Introduction}

Baryon-acoustic oscillations (BAOs) have emerged as one of the major probes of the expansion history of the Universe. In the early Universe, the ionized baryons were kinematically coupled to the cosmic microwave background (CMB) of photons via Thomson scattering. This baryon-photon fluid supported sound waves, sourced by primordial perturbations, that could travel a comoving distance $r_{\rm d}$ prior to decoupling. This distance is precisely constrained by CMB observations to be $r_{\rm d} = 147.3\pm 0.3$ Mpc \cite{planck15arxiv}. After decoupling, the baryons became effectively pressureless at large scales, and the perturbations in the baryons and dark matter grew together in a single combined growing mode. Thus at low redshift, {\em all} tracers of the matter density, either in dark matter or baryons, are predicted to show a feature in their correlation function at a position $r_{\rm d}$ -- or equivalently oscillatory features in their power spectrum $P(k)$, with spacing $\Delta k = 2\pi/r_{\rm d}$. This feature acts as a standard ruler, enabling galaxy redshift surveys to measure the distance-redshift relation $D(z)$, and (using the radial direction) the expansion rate $H(z)$. The BAO scale is of interest because its distinctive shape and large scale make it less dependent on nonlinear evolution and galaxy formation physics than the broad-band power spectrum \cite{albrecht06arxiv, seo08, padmanabhan09, seo10, mehta11}. However, the small amplitude of the feature makes it detectable only in very large surveys \cite{seo07}.

The early detections of the BAO in the clustering of low-redshift galaxies \cite{cole05, eisenstein05} have given way to a string of results of ever-increasing precision \cite{percival07, percival10, blake11, blake11a, padmanabhan12, xu12, kazin13, anderson14, kazin14, cuesta15arxiv}. High-redshift measurements have become possible by using quasar spectra to trace large-scale structure in the autocorrelation function of the Lyman-$\alpha$ forest and in its cross-correlation with quasars \cite{busca13, kirkby13, slosar13, font14, delubac15}. Taken together, these measurements have become one of the most important constraints on dark energy models \cite{aubourg15}. These successes have motivated a suite of future spectroscopic surveys to measure BAOs more precisely, including the Prime Focus Spectrograph \cite{takada14} and the Dark Energy Spectroscopic Instrument (DESI) \cite{levi13} in the optical, and {\slshape Euclid} \cite{laureijs11} and the {\slshape Wide-Field Infrared Survey Telescope} ({\slshape WFIRST}) \cite{spergel15arxiv} in the infrared. They have also spawned novel concepts for measuring BAOs such as radio intensity mapping \cite{wyithe08, chang08} as planned for e.g.\ the Canadian Hydrogen Intensity Mapping Experiment (CHIME) \cite{bandura14}.

The same acoustic oscillations that give rise to the BAO also leave the baryons with an r.m.s.\ velocity of $33$ km/s relative to the dark matter at decoupling, coherent over scales of many comoving Mpc. This velocity is cosmologically small at late times, since it decays $\propto 1/a$. However, it was realized in 2010 that the sound speed in neutral hydrogen at the decoupling epoch is only 6 km/s, so this ``streaming velocity'' is supersonic \cite{tseliakhovich10} and hence is more important than standard Jeans-like filtering in determining the scale on which baryons can fall into dark matter potential wells. The filtering mass of the cold IGM (before re-heating by astrophysical sources) is increased typically by a factor of $\sim 8$ relative to what it would be without the streaming velocities \cite{tseliakhovich11}. Moreover, the streaming velocity has order-unity spatial variations and a power spectrum showing prominent acoustic peaks \cite{tseliakhovich10}. The effects of streaming velocities on gas accretion and cooling have been a subject of intense analytical and numerical investigation \cite{stacy11, maio11, greif11, mcquinn12, fialkov12, naoz13, richardson13, asaba15arxiv}.

It was soon realized that these ingredients implied that small, high-redshift galaxies whose abundance was modulated by the streaming velocity would show an unusual BAO signature \cite{dalal10}, and in some models the BAO signature in the pre-reionization 21 cm signal could be strongly enhanced relative to the strength of the BAOs in the matter clustering alone \cite{visbal12, mcquinn12, fialkov14}. Moreover, if low-redshift galaxies have any memory of the streaming velocity, then low-redshift BAO measurements could be biased \cite{dalal10, yoo11}. While the direct effect of streaming velocities is on the small scale structure ($\lesssim\,$few$\,\times 10^6 M_\odot$), feedback processes associated with reionization or metal enrichment could influence the subsequent evolution of more massive galaxies in a way that is difficult to predict from first principles \cite{dalal10, yoo11}. In the absence of a first-principles theory, this effect can be parameterized in terms of the ``streaming velocity bias'' $b_v$, which is the excess probability to find a galaxy in a region with r.m.s.\ streaming velocity versus a region with zero streaming velocity. Studies based on perturbation theory have found that the BAO ruler shrinks (stretches) for $b_v>0$ ($b_v<0$) \cite{yoo11, yoo13, slepian15}.

In this paper, we compute the effect of streaming velocities on the BAO feature including all leading-order terms; we find that the largest term was missing from previous work. The galaxy density, a scalar, cannot depend on the direction of the streaming velocity, but only on its magnitude (or square). In linear perturbation theory with Gaussian initial conditions, the density (an odd moment) cannot correlate with the velocity-squared (an even moment), one must go to the next order to obtain a nonzero result. Previous investigations included three such effects: (i) the nonlinear evolution of the matter density field; (ii) nonlinear galaxy bias; and (iii) the autocorrelation of the streaming velocity field. We show that to consistent order in perturbation theory, two additional terms appear: (iv) the dependence of the galaxy abundance on the local tidal field \cite{baldauf12}; and (v) an ``advection term,'' since galaxy properties depend on the past streaming velocity at their Lagrangian position. We find that for plausible bias parameters, the tidal effect is small, but the advection term greatly enhances the shift in BAO position and impacts the shape and amplitude of the BAO feature. Because knowing the correct BAO scale is required to relate observed galaxy clustering to underlying cosmological physics, understanding the impact of streaming velocities is critical if we are to obtain unbiased results from the future generation of high-precision measurements.

\section{Formalism and bias model}
\label{sec:formalism}

We first construct a model for the distribution of galaxies. We describe our notations, choice of cosmology, and normalization conventions in \S\ref{ss:conventions}, before proceeding in \S\ref{ss:gbias} to building the model for the distribution of galaxies.

\subsection{Conventions}
\label{ss:conventions}

We work in real space in this paper, leaving the redshift-space treatment to future work. The fiducial cosmology is the base 6-parameter {\slshape Planck} + ``everything'' model \cite{planck15arxiv}: flat $\Lambda$CDM with $\Omega_{\rm b}h^2 = 0.02230$; $\Omega_{\rm m}h^2 = 0.14170$; $H_0=67.74$ km s$^{-1}$ Mpc$^{-1}$; $A_s=2.142\times 10^{-9}$ (at $k=0.05$ Mpc$^{-1}$); $n_s = 0.9667$; and $\tau = 0.066$.

The streaming velocity field $\mathbf v_{bc}\equiv\mathbf v_b-\mathbf v_c$ can be computed on large scales in linear perturbation theory, and scales $\propto 1/a$ once the baryons have decoupled and are effectively pressureless. Following the notation of Ref.~\cite{slepian15}, we define the normalized streaming velocity field to be
\begin{equation}
\label{eq:v_def}
\mathbf v_s(\mathbf x) = \frac{\mathbf v_{bc}(\mathbf x,a)}{\sigma_{vbc}}
= \frac{\mathbf v_{bc}(\mathbf x,a)}{\langle \mathbf v^2_{bc}(\mathbf x',a)\rangle_{\mathbf x'}^{1/2}},
\end{equation}
where the average in the denominator is taken over all positions $\mathbf x'$. By dividing out $\sigma_{vbc}$, we obtain a normalized streaming velocity that is independent of redshift and is of order unity. Note that some authors \cite{yoo11, yoo13} have defined an alternative variable $\mathbf u_r$, equivalent to $\sqrt3\, \mathbf v_s$ here, which has an r.m.s.\ value of 1 {\em per axis} (see Eq.~\ref{eq:dg-model}).

At linear order, the relative velocity field in Fourier space can be written as
\begin{equation}
\mathbf{v}_s(\mathbf{k}) = -i T_v(k,z)\hat{\mathbf{k}}\delta_{\rm lin}(\mathbf{k},z),
\end{equation}
where $T_v \propto (T_{v,b}-T_{v,c})/T_m$ is the ratio of transfer functions that map initial curvature fluctuations into late-time matter and velocity fluctuations. Note that with this definition, even though $\mathbf v_s(\mathbf k)$ is redshift-independent, $T_v$ decays as $\propto 1/D(z)$, where $D(z)$ is the growth factor. The appropriate transfer function can be obtained from a Boltzmann code -- we used both CAMB \cite{lewis00} and CLASS \cite{blas11}, obtaining consistent results -- and the normalization of Eq.~(\ref{eq:v_def}) at any desired redshift $z$ can be obtained by enforcing the integral:
\begin{equation}
\int_0^{k_{\rm max}} \frac{k^2P_{m,\rm lin}(k,z)}{2\pi^2}|T_v(k,z)|^2\, dk = 1,
\label{eq:normInt}
\end{equation}
where $P_{m,\rm lin}(k,z)$ is the linear matter power spectrum. The choice of $k_{\rm max}$ is set by the minimum scale relevant for the formation of the relevant tracer, e.g.\ its Lagrangian radius. In practice, we find that $\sigma_{vbc}$ is nearly insensitive to the choice of $k_{\rm max}$ unless fluctuations below the pre-reionization baryonic Jeans scale are included. These fluctuations are not relevant for galaxy formation, and we thus choose $k_{\rm max} = 10h\,{\rm Mpc}^{-1}$.

We will also need the tidal field magnitude $s^2 = s_{ij} s_{ij}$. Here the traceless-symmetric dimensionless tidal tensor $s_{ij}$ is given by
$s_{ij}(\mathbf{x}) = (\nabla_i \nabla_j \nabla^{-2} - \frac{1}{3}\delta_{ij}) \delta(\mathbf{x})$.

\subsection{Galaxy biasing model}
\label{ss:gbias}

We now write a model for the overdensity of a given tracer of large-scale structure, $\delta_g$. This tracer population could be e.g.\ galaxies, Lyman-$\alpha$ absorption, or the unresolved H {\sc i} 21 cm emissivity. While the detailed physics of the formation and evolution of these tracer populations remains an outstanding problem, nonlinear galaxy biasing \cite{mcdonald09b} provides a useful framework to study the streaming velocity effect. This theory is based on the idea that galaxy formation is {\em local}, with the only long-range physics being gravity. Under these assumptions, the galaxy overdensity measured on scales large compared to the range of galaxy formation physics should depend only on the density and tidal fields, the local streaming velocity, and their past history (since galaxy formation, while local in space, is obviously not local in time). At small scales, additional terms can appear involving derivatives of the density or tidal field, but as we are interested in large scales we do not include these. Note that any terms involving past history should be based on the history at fixed Lagrangian position, since small-scale structure, metal enrichment, and similar properties are advected by large-scale velocity fields rather than remaining in a fixed Eulerian cell.

Since galaxy overdensity is a scalar, its dependence on $\mathbf{v}_s$ must be at least quadratic. The leading corrections to the linear galaxy 2-point function are from terms of $\mathcal{O}(\dl^4)$ in the linear density field $\dl$, since terms of  $\mathcal{O}(\dl^3)$ vanish for Gaussian initial conditions. It follows that the galaxy biasing model we require should go up to $\mathcal{O}(\dl^3)$. Since our primary interest is the contributions coming from streaming velocities, we neglect $\mathcal{O}(\dl^3)$ contributions to the density field that do not involve $\mathbf{v}_{s}$ (e.g.\ \cite{saito14}). The tracer density is then given by:
\begin{align}
\dg(\mathbf{x}) =&\, b_1 \delta(\mathbf{x}) + \frac{b_2}{2} [\delta^2(\mathbf{x}) - \sigma^2 ] 
+ \frac{b_s}{2} [ s^2(\mathbf{x}) - \langle s^2 \rangle ] + \cdots
\nonumber \\ &
 + b_v [ v_s^2(\mathbf{q}) - 1 ] + b_{1v} \delta(\mathbf{x}) [v_s^2(\mathbf{q})-1 ]
 \nonumber \\ &
 + b_{sv}s_{ij}(\mathbf{x}) v_{s,i}(\mathbf{q}) v_{s,j}(\mathbf{q}) + \cdots,
\label{eq:dg-model}
\end{align}
where $\delta$ denotes the (nonlinear) dark matter density field and $\sigma^2$ is the variance in density fluctuations. The Lagrangian position (i.e. comoving position of the particles just after the Big Bang) is denoted $\mathbf{q}$ to distinguish it from Eulerian position $\mathbf{x}$. In this formulation, it is the linear $\mathbf{v}_s$ that is evaluated at $\mathbf{q}$, while the advection of the density and tidal fields is already included through the perturbative expansion of the density field. Although Eq.~(\ref{eq:dg-model}) expresses the astrophysical motivation for the advection contribution, this term can be derived from a purely Eulerian perspective, as shown in Appendix~\ref{app:Eulerian}. Indeed, as we demonstrate, the advection term is required to preserve Galilean invariance.

The definitions of the bias coefficients are not standardized: while our $b_v$ is equivalent to that of \cite{slepian15}, $b_r$ in \cite{yoo13} is related to both via $b_r=\frac{1}{3}b_v$. Note also that there are multiple combinatoric conventions for $b_2$.

The mapping between $\mathbf x$ and $\mathbf q$ can be expanded to order $\dl$, since we are only concerned with contributions to $\dg$ up to $\mathcal{O}(\dl^3)$. Lagrangian and Eulerian positions are related by $\mathbf{x}(\mathbf{q}, \eta)=\mathbf{q} + \mathbf{\Psi}(\mathbf{q}, \eta)$. The Lagrangian displacement is given to linear order by \mbox{$\mathbf{\Psi}=-\mathbf{\nabla}\nabla^{-2}\delta_\text{lin}(\mathbf{x},\eta)$} (this is the Zel'dovich approximation, combined with the fact that at leading order we do not need to distinguish $\mathbf x$ and $\mathbf q$ in the argument of a perturbation field).  Any field $\varphi$ then maps according to
\begin{equation}
\varphi(\mathbf{q}) =
\varphi(\mathbf{x}-\mathbf\Psi) =
 \varphi(\mathbf{x}) + \mathbf{\nabla} \varphi(x) \cdot \mathbf{\nabla} \nabla^{-2}  \delta_\text{lin} (\mathbf{x}, \eta) + \cdots.
\end{equation}
At the required order,
\begin{equation}
v_s^2(\mathbf{q})
= {v}_s^2(\mathbf{x}) + [\nabla_i \nabla^{-2}\delta_{\rm lin}(\mathbf{x})] [\nabla_i {v}_s^2(\mathbf{x})] .
\label{eq:mapping}
\end{equation}

\section{Effect on 2-point statistics and BAO position}
\label{sec:2-point}

We are interested in the tracer auto-correlation function $\xi_{\rm gg}(\mathbf{r}) = \langle \dg(\mathbf{x}) \dg(\mathbf{x'})\rangle$,
where ${\mathbf r}={\mathbf x}-{\mathbf x}'$. In this work, we consider terms that contribute at up to one-loop, i.e.\ $\mathcal{O}(\dl^4)$:
\begin{align}
\xi_{\rm gg}({\mathbf r}) =&\,\, b_1^2 \langle \delta | \delta\rangle + b_1 b_2 \langle \delta | \delta^2\rangle + b_1 b_s \langle \delta | s^2\rangle + \frac{1}{4}b_2^2 \langle \delta^2 | \delta^2\rangle
\nonumber \\
& + \frac{1}{4}b_s^2 \langle s^2 | s^2\rangle + \frac{1}{2}b_2 b_s \langle \delta^2 | s^2\rangle \nonumber \\
& + 2 b_1 b_v \left[ \langle \delta | \mathbf{v_s}^2 \rangle + \langle \delta | \nabla_i\nabla^{-2}\delta\nabla_i\mathbf{v_s}^2 \rangle \right]
\nonumber \\
& + b_{2}b_v \langle \delta^2 |\mathbf{v}_s^2\rangle + b_{s}b_v \langle s^2 | \mathbf{v}_s^2\rangle+ b_v^2 \langle \mathbf{v}_s^2 | \mathbf{v}_s^2 \rangle,
\label{eq:corr_funct}
\end{align}
where we use the shorthand $\langle A|B\rangle\equiv \langle A(\mathbf{x}) B(\mathbf{x}')\rangle$.

We note that there is no term proportional to $b_1 b_{1v}$ or $b_1 b_{sv}$, since by parity a scalar or tensor must have zero average correlation with a vector at the same position, and hence all contractions vanish when Wick's theorem is applied. In the following, we denote the streaming velocity correlations in Eq.~(\ref{eq:corr_funct}) as $\xi_{\delta v^2}$, $\xi_{\rm adv}$, $\xi_{\delta^2 v^2}$, $\xi_{s^2 v^2}$, and $\xi_{v^2 v^2}$, respectively. See Appendix~\ref{app:details} for the details of how all relevant correlations are calculated.

We use Wick's theorem to simplify the advection term:
\begin{align}
\xi_{\rm adv}({\mathbf r}) =&\,\langle \delta | \nabla_i\nabla^{-2}\delta\nabla_i\mathbf{v_s}^2 \rangle
\nonumber \\
=&\,
2\langle \delta_{\rm lin}({\mathbf x}) [\nabla_i \nabla^{-2}\delta_{\rm lin}(\mathbf{x}')] [{v}_{s,j}(\mathbf{x}') \nabla_i {v}_{s,j}(\mathbf{x}')]\rangle
\nonumber \\ =&\, 
\frac{2}{3}L_s \langle \dl (\mathbf{x}) \boldsymbol{\nabla} \cdot \mathbf{v}_s ( \mathbf{x}') \rangle ,
\label{eq:xi-adv.1}
\end{align} 
where
\begin{align}
\label{eqn:L_s:def}
L_s = \langle [ \boldsymbol{\nabla} \nabla^{-2} \dl ( \mathbf{x}) ] \cdot \mathbf{v}_s( \mathbf{x}) \rangle ,
\end{align} 
and we note that only one of the three contractions in the second line of Eq.~(\ref{eq:xi-adv.1}) is nonzero (the others vanish since by isotropy the symmetric tensor $\nabla_i {v}_{s,j}(\mathbf{x}')$ has zero correlation with the vectors $\nabla_i \nabla^{-2}\delta_{\rm lin}(\mathbf{x}')$ or ${v}_{s,j}(\mathbf{x}')$ at the same point).

To illustrate the impact of streaming velocities and this new advection term, we show results for a fiducial sample of emission line galaxies (ELGs) at $z=1.2$, such as that relevant for DESI, {\slshape Euclid}, and {\slshape WFIRST}. Unless otherwise noted, we assume $b_1=1.5$, $b_2=0.25$, and $b_s=\frac27(1-b_1)=-0.14$. However, the impact of streaming velocities depends primarily on the ratio $b_v/b_1$ for the tracer in question, and thus our results qualitatively hold for other samples.

Due to nonlinear evolution, the BAO in the dark matter correlation $\langle \delta | \delta\rangle$ is shifted from its linear position. To model this, we include the one-loop standard perturbation theory (SPT) contributions to the matter power spectrum \cite{bernardeau02}. As can be seen in Fig.~\ref{fig:BAO_shift}, these terms lead to a $\sim0.2\%$ shift in the BAO at $z=1.2$. We note that SPT does not provide the ideal model for the evolved BAO -- we leave a more detailed treatment of this effect for future work. The inclusion of these nonlinear terms alters the impact of streaming velocities when fitting the BAO position -- nonlinear broadening makes the BAO feature more sensitive to the shift from streaming velocities. Note that Ref.~\cite{yoo13} modeled the nonlinear matter power spectrum using Halofit \cite{smith03}, which does not include nonlinear evolution of the BAO (see their Fig.~3).

\begin{figure}
\centering
\includegraphics[width=\columnwidth]{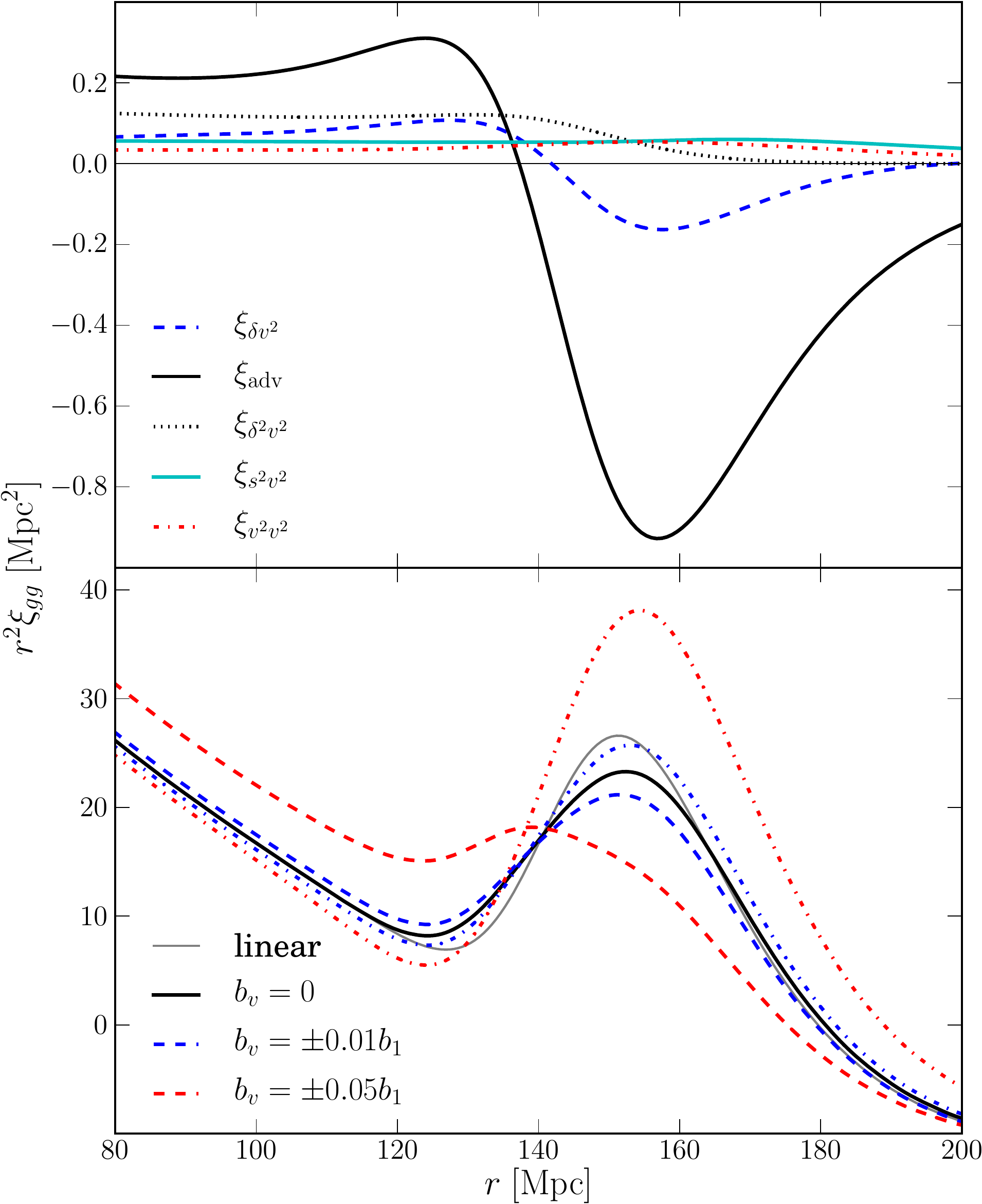}
\caption{{\it Top panel:} All contributions to the correlation function from streaming velocities up to $\mathcal{O}(\dl^4)$ are shown at $z=1.2$, with $b_1=b_2=b_s=1$, and $b_v=0.01$. The new advection term (black solid line) is the dominant effect. {\it Bottom panel:} The ELG correlation function is shown for fiducial bias values ($b_1=1.5,~b_2=0.25,~b_s=-0.14$) at $z=1.2$ with different values of $b_v$. Dashed (dot-dashed) indicates positive (negative) $b_v$. For reference, the thin solid (grey) line shows the linear theory prediction.}
\label{fig:corr_comp}
\end{figure}

\begin{figure}
\centering
\includegraphics[width=\columnwidth]{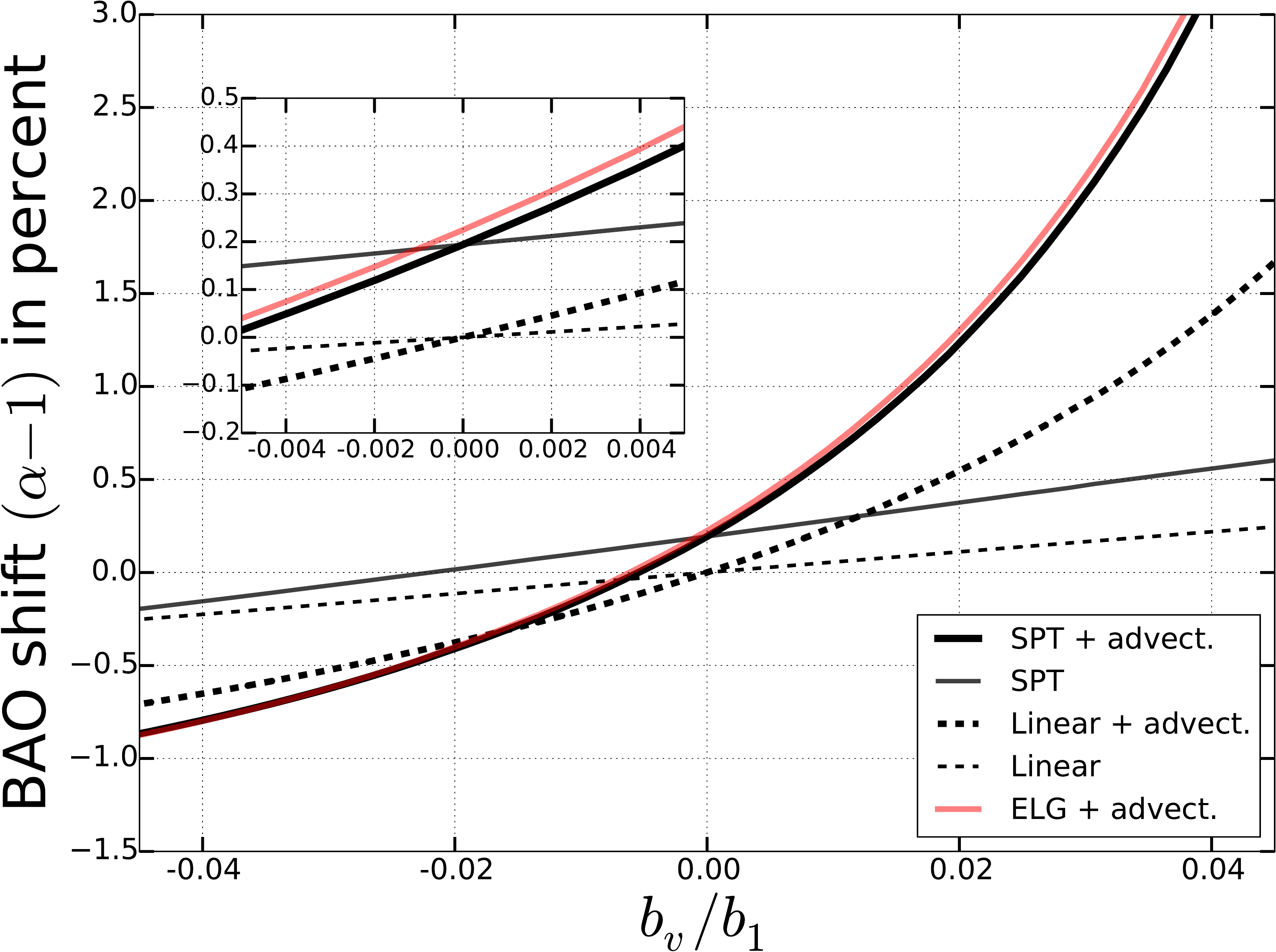}
\caption{The shift in BAO position due to streaming velocities is shown as a function of $b_v/b_1$. Thick (thin) lines show the shift with (without) the advection term. The solid black lines include the one-loop SPT correction to the dark matter power spectrum ($b_2=b_s=0$), while the solid grey line also includes fiducial $b_2$ and $b_s$ values for the ELG sample. Inset shows detailed behavior for small $b_v$.}
\label{fig:BAO_shift}
\end{figure}

Streaming velocity contributions to the correlation function (including all prefactors) are plotted in the top panel of Fig.~\ref{fig:corr_comp}. For reasonable bias values, $\xi_{\delta v^2}$ had been considered the primary streaming velocity term. The new advection effect is larger by a factor of $\sim 5$. The bottom panel of Fig.~\ref{fig:corr_comp} shows the ELG correlation function with different values of $b_v$ -- the impact on both the shape and position of the BAO feature is apparent.

To quantify the shift of the BAO peak due to relative velocity effects, we employ a method similar to \cite{seo08, yoo13}, fitting the shifted power spectrum to a template with flexible broadband power -- see Appendix~\ref{app:BAOshift} for more details. Figure \ref{fig:BAO_shift} shows the BAO shift as a function of $b_v/b_1$, both including and ignoring contributions from nonlinear galaxy bias and BAO evolution. For positive $b_v/b_1$ streaming velocities damp the BAO feature and shift it to smaller scales. For negative $b_v/b_1$, streaming velocities enhance and quickly dominate the BAO feature as $|b_v|$ increases, leading to a saturation in the effective shift. Note that we differ from Ref.~\cite{yoo13} by an overall factor of 2 in the numerical evaluation of $\mathcal{O}(\dl^4)$ terms and find a correspondingly smaller shift in the BAO position from the non-advection terms they consider.

\section{Conclusions}
\label{sec:conc}

We have examined the impact of streaming velocities on the tracer correlation function, considering all contributions at $\mathcal{O}(\dl^4)$ and including two terms not considered in previous work. While we find the correlation of the tidal field and the streaming velocity to be small, the contribution from advection is significant, dominating the total effect of streaming velocities on the BAO feature. The importance of advection is due to the rapid change in streaming velocity correlations at the BAO scale.  For a simple illustration, consider a single $\delta$-function overdensity that has evolved to decoupling ($z \approx 1020$). Dark matter at all separations infalls towards the overdensity. Within the acoustic scale, baryons are roughly in hydrostatic equilibrium. Just inside the acoustic scale, baryons move outward due to radiation pressure, while just outside this scale, baryons match the dark matter infall (e.g.\ Figure 2 of Ref.~\cite{slepian15}). Thus, the streaming velocity, $v_{bc}$, rapidly changes at the acoustic scale, and advection can move tracers separated by roughly this scale between regions of different $v_{bc}$. Indeed, this effect is nearly maximal, since the first-order displacement is almost entirely anti-correlated with the relative velocity direction (correlation coefficient of $\sim -0.9$). The qualitative behavior expected from this simplified picture can be seen in Figure~\ref{fig:corr_comp}: at the BAO scale, advection has carried in tracers that formed at slightly larger scales, where $v_{bc}$ is much smaller. Thus, for positive (negative) $b_v/b_1$ the observed correlation function is suppressed (enhanced). The overall effect is to shift the observed BAO feature to smaller (larger) scales and to suppress (enhance) its amplitude.

The effect of advection boosts the impact of $b_v$, dramatically increasing the range of parameter space over which streaming velocities are relevant to large-scale structure surveys. Conversely, advection makes $b_v$ significantly easier to detect, providing a potential window into the astrophysics of streaming velocities and tracer formation. For instance, DESI will obtain an overall BAO-scale measurement of order $0.2\%$ ($1\sigma$), corresponding to the shift induced by streaming velocities at $b_v/b_1 \sim 0.004$ \cite{levi13}. The ultimate impact of streaming velocities will depend on the as-yet-unknown value and sign of $b_v$, as well as other possible bias terms related to differences in the baryon and CDM fluids (see Appendix~\ref{app:Eulerian}). Their direct effect is to suppress the infall of baryons into halos by a fractional amount $\sim(10^5 M_\odot/M_{\rm halo})^{2/3}$ (e.g.\ \cite{tseliakhovich11}). This scaling results from the fact that the suppression is proportional to ${\mathbf v}_s^2$, and is ${\cal O}(1)$ when the streaming-enhanced filtering scale is the halo mass, which suggests a contribution to $b_v$ of order a few $\times 10^{-5}$ at galaxy mass scales. We view this as a ``soft'' lower bound on $b_v$ in the sense that e.g.\ reionization physics may be much more important and thus dominate $b_v$, but there is no reason for a precise cancellation that would give a total $b_v\approx 0$. On the other hand, very large values ($|b_v|/b_1 \gtrsim 0.1$) would disrupt the qualitative agreement with current observations. The effect of streaming velocities may also be relevant for other luminous tracers of large-scale structure, notably the Lyman-$\alpha$ forest and (possibly) unresolved 21 cm emission; these tracers are sensitive to a range of mass scales down to the post-reionization Jeans scale (few$\times 10^9$ $M_\odot$), and their $b_v$ may be correspondingly larger. We leave a more detailed consideration of astrophysical effects that impact $b_v$ for future work. We also defer consideration of reconstruction (which may have significant impact on displacements) \cite{eisenstein07b,padmanabhan09b} and redshift-space distortions in the context of streaming velocities.

While we have primarily considered the impact of streaming velocities on the {\em position} of the BAO feature, it is clear from Fig.~\ref{fig:corr_comp} that the BAO {\em shape} is also significantly altered. Although it is not typically not considered in cosmological analyses, these results suggest that the shape may help to separate the effect of streaming velocities from geometric effects. We will consider implications of changes to the BAO shape in future work.

\acknowledgements

We thank John Beacom, Florian Beutler, Peter Melchior, Ashley Ross, Uro\v{s} Seljak, Zachary Slepian, David Weinberg, and Jaiyul Yoo for helpful discussions. We are also grateful for suggestions from an anonymous referee which improved the paper.
J.B. is supported by a CCAPP fellowship.
J.M. is supported through NSF Grant AST-1516997.
C.H. is supported by the David and Lucile Packard Foundation, the Simons Foundation, and the U.S. Department of Energy.


\vspace{-0.3cm}
\appendix

\section{Details of calculations}
\label{app:details}

It is convenient to work calculations in Fourier space.  We use the relationship between fields in configuration and Fourier space:
$f({\mathbf k}) = \int d^3\mathbf x\,e^{-i\mathbf k\cdot\mathbf x}\, f(\mathbf x)$.
We define the power spectrum in terms of the ensemble average over the Fourier space density fields:
\begin{equation}
\langle \delta(\mathbf{k}) \delta(\mathbf{k'})\rangle = (2\pi)^3\delta^{(3)}(\mathbf{k}+\mathbf{k'})P(k).
\end{equation}
In perturbation theory we write the matter field as a series expansion: 
$\delta(\mathbf{k})= \delta^{(1)}(\mathbf{k}) + \delta^{(2)}(\mathbf{k}) + \cdots$,
in which terms of order 2 and higher represent non-linear evolution of the matter field. The second order contribution to the density contrast is 
\begin{align}
\delta^{(2)}(\mathbf{k}) = \int \frac{ d^3 \mathbf{k}_1}{(2 \pi)^3} F_2( \mathbf{k}_1, \mathbf{k}_2) \delta^{(1)}(\mathbf{k}_1) \delta^{(1)}(\mathbf{k}_2) , 
\end{align} 
where we define
$\mathbf{k}_2 \equiv \mathbf{k}-\mathbf{k}_1$, $\mu_{12} \equiv \mathbf{k}_1\cdot\mathbf{k}_2/(k_1k_2)$, and
the second order density kernel is 
\begin{align}
F_2(\mathbf{k}_1, \mathbf{k}_2) = \frac{5}{7} + \frac{1}{2}\mu_{12}\left( \frac{k_1}{k_2}+\frac{k_2}{k_1} \right) + \frac27\mu_{12}^2.
\end{align}
In Fourier space the squared tidal tensor is 
\begin{align} 
s^2(\mathbf{k})= \int \frac{d ^3 \mathbf{k}_1 }{(2 \pi)^3} S_2( \mathbf{k}_1, \mathbf{k}_2 ) \delta(\mathbf{k}_1) \delta(\mathbf{k}_2) , 
\end{align}
where $S_2(\mathbf{k}_1, \mathbf{k}_2) = \mu_{12}^2 - \frac{1}{3}$.
At one-loop, the correlations in Eq.~(\ref{eq:corr_funct}) can be expressed as Fourier transforms of the corresponding power spectra:
\begin{align}
\label{eqn:pks}
P_{\rm adv}(k) =\, & \frac{2}{3} L_s k T_v(k) \Pl(k),
\nonumber \\
P_{\delta \delta^{2}}(k) =\,& 2 \int  \frac{d ^3 \mathbf{k}_1}{(2 \pi)^3}  F_2( \mathbf{k}_1,  \mathbf{k}_2 ) \Pl(k_1) \Pl(k_2)  , 
\nonumber \\
P_{\delta^{2} \delta^{2}}(k) =\,& 2 \int  \frac{d ^3 \mathbf{k}_1 }{(2 \pi)^3} \Pl(k_1) \Pl(k_2 )  , 
\nonumber \\
P_{\delta v^2}(k) =\,& -2 \int  \frac{d ^3 \mathbf{k_1}}{(2 \pi)^3}  F_2( \mathbf{k}_1,  \mathbf{k}_2 ) \mu_{12} T_v(k_1) T_v(k_2) \nonumber \\ & \times
\Pl(k_1) \Pl(k_2) , 
\nonumber \\
P_{\delta^{2} v^2}(k) =\,& -2 \int \frac{d ^3 \mathbf{k}_1}{(2 \pi)^3} \mu_{12} T_v(k_1) T_v(k_2) \Pl(k_1) \Pl(k_2 ) , 
\nonumber \\
P_{s^{2} v^2}(k) =\,& -2 \int \frac{d ^3 \mathbf{k}_1}{(2 \pi)^3} \mu_{12} S_2( \mathbf{k}_1,  \mathbf{k}_2 )T_v(k_1) T_v(k_2) \nonumber \\ & \times
\Pl(k_1) \Pl(k_2 ) , 
\nonumber \\
P_{v^{2} v^2}(k) =\,& 2 \int \frac{d ^3 \mathbf{k}_1}{(2 \pi)^3}  \mu_{12}^2 T^2_v(k_1) T^2_v(k_2) \Pl(k_1) \Pl(k_2) , 
\nonumber \\
P_{\delta^{2} s^2}(k) =\,& 2 \int \frac{d ^3 \mathbf{k}_1}{(2 \pi)^3} S_2( \mathbf{k}_1,  \mathbf{k}_2 ) \Pl(k_1) \Pl( k_2 ) , 
\nonumber \\
P_{s^{2} s^2}(k) =\,& 2 \int \frac{d ^3 \mathbf{k}_1}{(2 \pi)^3} S^2_2( \mathbf{k}_1,  \mathbf{k}_2 ) \Pl(k_1) \Pl( k_2 ) ,~{\rm and} 
\nonumber \\
P_{\delta s^2}(k) =\,& 2 \int \frac{d ^3 \mathbf{k}_1}{(2 \pi)^3} F_2( \mathbf{k}_1,  \mathbf{k}_2  )S_2( \mathbf{k}_1,  \mathbf{k}_2 ) \nonumber \\ & \times
\Pl(k_1) \Pl( k_2 ) .
\end{align} 
Here the Fourier-space expression for $L_s$, defined in Eq.~(\ref{eqn:L_s:def}), is
\begin{align} 
\label{eqn:L_s}
L_s=\int \frac{d ^3 \mathbf{k}}{(2 \pi)^3} \frac{T_v(k)}{k} \Pl(k) .
\end{align} 
Note that $L_s$ can be interpreted as the r.m.s.\ contribution to the comoving displacement that is correlated with the streaming velocity (opposite directions) at the same position. At $z=1.2$, $L_s=7.7~\rm{Mpc}$, compared with an r.m.s.\ displacement of $\langle\Psi^2\rangle^{1/2} = 8.5~\rm{Mpc}$. These power spectra must be multiplied by the relevant numerical pre-factors, including bias values, shown in Eq.~(\ref{eq:corr_funct}). The full galaxy power spectrum is then
\begin{align} \label{P_g} P_g(k)=&\, b_1^2 P_{\rm NL}(k) + b_1b_2 P_{ \delta \delta^{2}}(k) + \frac{1}{4} b^2_2 P_{ \delta^{2} \delta^{2}}(k)
\nonumber \\ &
+  \frac{1}{4} b^2_s P_{s^2 s^2}(k)  + b_1 b_s P_{ \delta s^2}(k) + \frac{1}{2} b_2b_s P_{ \delta^{2} s^2}(k)
\nonumber \\
& + 2 b_1 b_v \Bigl[ P_{ \delta v^2}(k) + P_{\rm adv}(k) \Bigr]
\nonumber \\
&+ \frac{1}{2}b_2 b_v P_{ \delta^{2} v^2}(k) + \frac{1}{2} b_s b_v P_{ s^2 v^2}(k) + b_v^2 P_{ v^2 v^2} (k),
\end{align}
where $P_{\rm NL}(k) = P_{11}(k) + P_{22}(k) + P_{13}(k)$ is the non-linear power spectrum calculated in standard perturbation theory to one-loop order.

\section{Calculating the BAO shift}
\label{app:BAOshift}

Following \cite{seo08, yoo13}, we use the following template power spectrum: 
\begin{align}
P_{\rm model}(k)= \sum_{j=0}^2 c_jk^j P_{\rm evo}(k/\alpha) + \sum_{j=0}^5 a_jk^j,
\end{align} 
where the coefficients $a_i$ and $c_i$ are marginalized as nuisance parameters and $\alpha$ represents the shift in the BAO peak ($\alpha>1$ corresponds to a shift towards smaller scales). Our fitting template differs slightly from \cite{seo08, yoo13} in that we omit the $a_7 k^7$ term. We found that including this term provided too much flexibility, leading to persistent residual noise in the fit, although results were otherwise nearly identical. The nonlinear damping of the BAO is captured by the evolved power spectrum:
\begin{align}P_{\rm evo} (k) = \left[ \Pl(k) - P_{\rm nw}(k) \right] e^{-k^2 \Sigma_m^2/2}  + P_{\rm nw} (k) , 
\end{align} 
where $P_{\rm nw}$ is the no-wiggle (i.e.\ BAO-removed) power spectrum, computed from the fitting formula of \cite{eisenstein98}, and $\Sigma_m=4.8\hMpc$ is the fiducial damping factor at $z=1.2$, although it is treated as a free parameter in our analysis. While this template includes nonlinear damping of the BAO, it does not include the corresponding shift in BAO position ($\sim0.2\%$ when $b_v=0$).

The power spectrum in Eq.~\ref{P_g} (corresponding to the Fourier-space analog of Eq.~\ref{eq:corr_funct}) is fit to the template by minimizing $\chi^2$ assuming the standard covariance for $P(k)$ (e.g.\ \cite{tegmark97}): 
\begin{align}
\chi^2 = V \int_{k_{\rm min}}^{k_{\rm max}}  \frac{d^3k }{(2 \pi)^3} \frac{ [ P_g(k) - P_{\rm model}(k)]^2}{2[ P_g(k) + 1/\bar{n} ] ^2 } ~, 
\end{align} 
where we have chosen the density of galaxies $\bar{n}=3\times10^{-4}h^3\,{\rm Mpc}^{-3}$, and the integration is from $0.02<k<0.35h\,\text{Mpc}^{-1}$.

\section{Eulerian treatment of streaming velocities}
\label{app:Eulerian}

Although the physics of galaxy formation should be affected by the streaming velocity at the Lagrangian positions of tracers, these velocities can be equivalently expressed at the Eulerian positions. Indeed, as long as all relevant terms are included, there is always a mapping between Eulerian and Lagrangian biasing (e.g.\ \cite{baldauf12}). In this appendix, we demonstrate how consistent Eulerian treatment produces the same advection contribution to the streaming velocity bias.

We can write the streaming velocity as the Eulerian expansion: \mbox{$\mathbf v_{bc}(\mathbf{x})= \mathbf v_{bc}^{(1)}(\mathbf{x}) + \mathbf v_{bc}^{(2)}(\mathbf{x}) + \cdots$.} This expansion is analogous to the Eulerian treatment of the density field, and the higher-order streaming velocity contributions can be derived in similar fashion (e.g.\ \cite{jain94}). We start with the continuity and Euler equations for two fluids: cold dark matter and baryons. We assume curl-free velocity fields, which can be expressed in terms of the velocity divergence: $\theta(\mathbf{k}) = i \mathbf{k}\cdot\mathbf{v}(\mathbf{k})$. Working in an Einstein-de Sitter universe, the Fourier space equations are:
\begin{align}
\label{eqn:fluids}
& a \mathcal{H}(a)\frac{\partial\delta_x(\mathbf{k},a)}{\partial a} + \theta_x(\mathbf{k},a) \nonumber \\
& = - \int \frac{ d^3 \mathbf{k}_1}{(2\pi)^3} \frac{\mathbf{k}\cdot\mathbf{k}_1}{k_1^2}\theta_x(\mathbf{k}_1,a)\delta_x(\mathbf{k}_2,a) ~~{\rm and} \\
& \mathcal{H}(a) \left[a\frac{\partial\theta_x(\mathbf{k},a)}{\partial a} + \theta_x(\mathbf{k},a) + \frac{3}{2}\mathcal{H}(a) \delta_m(\mathbf{k},a) \right] \nonumber \\
& = - \int\frac{d^3 \mathbf{k}_1}{(2\pi)^3} \frac{k^2\mathbf{k}_1\cdot\mathbf{k}_2}{2k_1^2k_2^2}\theta_x(\mathbf{k}_1,a)\theta_x(\mathbf{k}_2,a) ,
\end{align}
where ``$x$'' denotes the relevant fluid (baryons or CDM), $\mathcal{H}(a) = a H(a)$ is the conformal Hubble parameter, and $\delta_m$ is the total matter density perturbation. Since we are interested in the {\it relative} velocity between dark matter and baryons, we express these equations in terms of total and relative densities and velocity divergences, defined by:
\begin{align}
\delta_m &= f_c \delta_c + f_b \delta_b ; ~ \theta_m = f_c \theta_c + f_b \theta_b \nonumber\\
\delta_{bc} &= \delta_b - \delta_c ; ~ \theta_{bc} = \theta_b - \theta_c,
\end{align}
where $f_c$ and $f_b$ are the CDM and baryon fractions, respectively, of the total matter, and $f_c+f_b=1$. The evolution of the total and relative quantities is then given by:
\begin{widetext}
\begin{align}
& a \mathcal{H}(a)\frac{\partial\delta_m(\mathbf{k},a)}{\partial a} + \theta_m(\mathbf{k},a)
 = - \int \frac{d^3 \mathbf{k}_1}{(2\pi)^3} \frac{\mathbf{k}\cdot\mathbf{k}_1}{k_1^2}\big\{\theta_m(\mathbf{k}_1,a)\delta_m(\mathbf{k}_2,a) + f_b f_c \theta_{bc}(\mathbf{k}_1,a)\delta_{bc}(\mathbf{k}_2,a) \big\} , \\
& \mathcal{H}(a) \left[a\frac{\partial\theta_m(\mathbf{k},a)}{\partial a} + \theta_m(\mathbf{k},a) + \frac{3}{2}\mathcal{H}(a)\delta_m(\mathbf{k},a) \right]
 = \! - \! \int \! \frac{ d^3 \mathbf{k}_1}{(2\pi)^3} \frac{k^2\mathbf{k}_1\cdot\mathbf{k}_2}{2k_1^2k_2^2} \big\{\theta_m(\mathbf{k}_1,a)\theta_m(\mathbf{k}_2,a) + f_b f_c \theta_{bc}(\mathbf{k}_1,a)\theta_{bc}(\mathbf{k}_2,a)\big\}, \\
\label{eqn:relcont}
& a \mathcal{H}(a)\frac{\partial\delta_{bc}(\mathbf{k},a)}{\partial a} + \theta_{bc}(\mathbf{k},a)
 = - \int \frac{ d^3 \mathbf{k}_1}{(2\pi)^3} \frac{\mathbf{k}\cdot\mathbf{k}_1}{k_1^2}\big\{[(f_c-f_b)\theta_{bc}(\mathbf{k}_1,a)\delta_{bc}(\mathbf{k}_2,a) \nonumber \\
 &~~~~~~~~~~~~~~~~~~~~~~~~~~~~~~~~~~~~~~~~~~~~~~~~~~~~~~~~~~~~~~~
 + \left[\theta_{bc}(\mathbf{k}_1,a)\delta_m(\mathbf{k}_2,a) + \theta_m(\mathbf{k}_1,a)\delta_{bc}(\mathbf{k}_2,a)\right] \big\} , \\
\label{eqn:releuler}
& \mathcal{H}(a) \left[a\frac{\partial\theta_{bc}(\mathbf{k},a)}{\partial a} + \theta_{bc}(\mathbf{k},a) \right]
 = - \int \frac{ d^3 \mathbf{k}_1}{(2\pi)^3} \frac{k^2\mathbf{k}_1\cdot\mathbf{k}_2}{2k_1^2k_2^2}\big\{(f_c - f_b)\theta_{bc}(\mathbf{k}_1,a)\theta_{bc}(\mathbf{k}_2,a) +  2\theta_{bc}(\mathbf{k}_1,a)\theta_m(\mathbf{k}_2,a)\big\}.
\end{align}
\clearpage
\end{widetext}
Setting the right hand sides of these equations to zero, we can solve for the linear evolution. The (standard) growing-mode solution for total matter fluctuations is:
\begin{equation}
\delta_m^{(1)}(\mathbf{k},a) \propto a  ~{\rm and}~
 \theta_m^{(1)}(\mathbf{k},a) = - \mathcal{H}(a)  \delta_m^{(1)}(\mathbf{k},a) \propto a^{1/2}.
\end{equation}
The relative velocity equation is a first-order differential equation, and its solution evolves as $\theta_{bc}^{(1)}(\mathbf{k},a)\propto a^{-1}$. The other two modes are the decaying matter density mode, $\delta_m^{(1)}(\mathbf{k},a) \propto a^{-3/2}$, and a mode with $\delta_{bc}^{(1)}(\mathbf{k},a)$ constant and zero velocities. We are interested in the nonlinear interaction when the growing density mode {\em and} the relative velocity mode are present in the initial conditions, not just the growing density mode.

We want the second-order relative velocity, $\theta_{bc}^{(2)}(\mathbf{k},a)$. This has contributions from both ``total-relative'' terms ($\theta_{bc}\theta_m$) and ``relative-relative'' terms ($\theta_{bc}\theta_{bc}$). Since the total and relative perturbations must be of similar magnitude at recombination, but relative velocities decline as $\propto a^{-1}$ whereas total velocities grow as $\propto a^{1/2}$, by $z \sim1$ the relative-relative coupling terms are suppressed by a factor of $\sim 10^4$ relative to the total-relative terms and can thus be safely ignored.

The equation for the second-order relative velocity is then
\begin{equation}
\frac{\partial[a\theta_{bc}^{(2)}(\mathbf{k},a)]}{\partial a} 
 = \int \frac{ d^3 \mathbf{k}_1}{(2\pi)^3} \frac{k^2\mathbf{k}_1\cdot\mathbf{k}_2}{k_1^2k_2^2} \theta_{bc}^{(1)}(\mathbf{k}_1,a)\delta_m^{(1)}(\mathbf{k}_2,a).
\end{equation}
The right-hand side is independent of $a$, so the solution is
\begin{equation}
a\theta_{bc}^{(2)}(\mathbf{k},a) = \left( a - a_{\rm i} \right) \int \frac{ d^3 \mathbf{k}_1}{(2\pi)^3} \frac{k^2\mathbf{k}_1\cdot\mathbf{k}_2}{k_1^2k_2^2} \theta_{bc}^{(1)}(\mathbf{k}_1)\delta_m^{(1)}(\mathbf{k}_2),
\end{equation}
where $a_{\rm i}$ is the initial scale factor, i.e. where $\theta_{bc}^{(2)}(\mathbf{k},a)=0$, which we take to be the time of recombination. At redshifts relevant for BAO measurements, $a_{\rm i}/a\ll 1$, and so
\begin{equation}
\theta_{bc}^{(2)}(\mathbf{k},a) = \int \frac{ d^3 \mathbf{k}_1}{(2\pi)^3} \frac{k^2\mathbf{k}_1\cdot\mathbf{k}_2}{k_1^2k_2^2} \theta_{bc}^{(1)}(\mathbf{k}_1)\delta_m^{(1)}(\mathbf{k}_2).
\end{equation}
Alternatively, we could make the following power-law ansatz for the leading time-dependent term at each order (e.g.\ \cite{goroff86}):
\begin{align}
\theta_{bc}^{(n)}(\mathbf{k},a) &=\mathcal{H}(a)a^{n-3/2}\theta_{bc}^{(n)}(\mathbf{k}), \nonumber \\
 \delta_{bc}^{(n)}(\mathbf{k},a) &= a^{n-3/2}\delta_{bc}^{(n)}(\mathbf{k}).
\end{align}
Requiring Eq.~(\ref{eqn:releuler}) to hold at each order and dropping the ``relative-relative'' terms as before yields:
\begin{align}
(n-1)\theta_{bc}^{(n)}(\mathbf{k}) =& -\int \frac{ d^3 \mathbf{k}_1}{(2\pi)^3} \frac{k^2\mathbf{k}_1\cdot\mathbf{k}_2}{k_1^2k_2^2} \nonumber \\
&\times \sum_{l=1}^{n-1} \theta_{bc}^{(l)}(\mathbf{k}_1)\theta_m^{(n-l)}(\mathbf{k}_2) .
\label{eqn:reldivalt}
\end{align}
For $n=2$, we recover the solution above.

Because its linear mode decays as $a^{-1}$, it is somewhat counterintuitive that $\mathbf{v}_{bc}$ has a significant nonlinear contribution. However, because it couples to the total matter perturbation, there are nonlinear corrections which grow in fractional importance as the matter fluctuations grow. The ratio of second- to first-order velocity perturbations $\theta_{bc}^{(2)}(\mathbf{k},a)/\theta_{bc}^{(1)}(\mathbf{k},a)$ scales as $\propto a$, i.e.\ in proportion to the growth factor of the growing matter mode.

Our next step is to determine the galaxy biasing terms that arise from the second-order streaming velocity. Using the definition of the velocity divergence and normalizing by the r.m.s. streaming velocity, the second-order streaming velocity is:
\begin{align}
\mathbf{v}_{s}^{(2)}(\mathbf{k}) = \mathbf{k} \int \frac{ d^3 \mathbf{k}_1}{(2\pi)^3} \frac{\mathbf{k}_1\cdot\mathbf{k}_2}{k_1^2k_2^2}\mathbf{k}_1\cdot\mathbf{v}_s^{(1)}(\mathbf{k}_1)\delta_m^{(1)}(\mathbf{k}_2).
\end{align}
This expression can be written in configuration space (using index notation):
\begin{align}
{v}_{s,i}^{(2)}(\mathbf{x}) &=
\nabla_i \left[ (\nabla_j\nabla^{-2} \nabla_k v_{s,k}^{(1)})( \nabla_j \nabla^{-2} \delta_m^{(1)}) \right]
\nonumber \\
 &= [\nabla_i v_{s,j}^{(1)}] \nabla_j\nabla^{-2}\delta_m^{(1)} + v_{s,j}^{(1)}\nabla_i\nabla_j\nabla^{-2}\delta_m^{(1)},
\end{align}
where the irrotationality of $\mathbf v_s^{(1)}$ was used to show that $\nabla_j\nabla^{-2} \nabla_k v_{s,k}^{(1)} = v_{s,j}^{(1)}$, and 
where all quantities are evaluated at Eulerian position $\mathbf{x}$. Finally, we can write $v_s^2(\mathbf{x})$ up to $\mathcal{O}(\dl^3)$:
\begin{align}
v_s^2(\mathbf{x}) =& \mathbf{v}_s^{(1)} \cdot \mathbf{v}_s^{(1)} + 2 \mathbf{v}_s^{(1)} \cdot \mathbf{v}_s^{(2)} ~~~~~~~~\nonumber\\
=& [v_s^{(1)}]^2 + 2v_{s,i}^{(1)}[\nabla_i v_{s,j}^{(1)}] \nabla_j\nabla^{-2}\delta_m^{(1)}
\nonumber \\
& + 2v_{s,i}^{(1)} v_{s,j}^{(1)}\nabla_i\nabla_j\nabla^{-2}\delta_m^{(1)}
\nonumber \\
=& [v_s^{(1)}]^2 + 2v_{s,i}^{(1)}[\nabla_j v_{s,i}^{(1)}] \nabla_j\nabla^{-2}\delta_m^{(1)}
\nonumber \\
& + 2v_{s,i}^{(1)} v_{s,j}^{(1)}\left[s_{ij}^{(1)}+ \frac13\delta_m^{(1)} \right]
\nonumber \\
=& [v_s^{(1)}]^2 + \boldsymbol{\nabla}\{[v_s^{(1)}]^2\}\cdot [ \boldsymbol{\nabla}\nabla^{-2}\delta_m^{(1)} ]
\nonumber\\
&+ 2v_{s,i}^{(1)}v_{s,j}^{(1)}s_{ij}^{(1)}+ \frac{2}{3}{v_s^{(1)}}^2\delta_m^{(1)}.
\end{align}
(The irrotationality of $\mathbf v_s^{(1)}$ was used again to swap the indices in $\nabla_i v_{s,j}^{(1)}$.) The second term is the advection contribution (see Eq.~\ref{eq:mapping}), while the final two terms can be absorbed into the definition of $b_{1v}$ and $b_{sv}$ in Eq.~(\ref{eq:dg-model}), indicating that they would naturally appear even had we not originally included them.

In the context of separate baryon and CDM fluids, symmetry allows additional biasing terms, proportional to $\delta_{bc}$ or $\theta_{bc}$. Most notably, a $\theta_{bc}$ contribution would correlate with the linear bias to produce a term with the same scale dependence as $\xi_{\rm adv}$. Such a bias contribution would thus be partially degenerate with $b_v$, although they each additionally produce unique correlations. Although such bias terms may may be physically motivated, they are distinct effects and can be self-consistently set to zero in our present treatment. The dominant contributions (in terms of redshift scaling) to $\mathbf{v}_{bc}^2$ at each perturbative order will not contain a linear contribution from $\delta_{bc}$ (see Eqs.~\ref{eqn:reldivalt} and \ref{eqn:releuler}), while $\theta_{bc}$ is the derivative of $v_{bc}$ and is thus suppressed by a factor of $k$. We leave consideration of these terms for future work.

Finally, we show that the advection contribution is required to maintain Galilean invariance, which was actually broken in previous presentations of streaming velocity bias. Galilean invariance requires that the laws of physics (e.g.\ galaxy formation) are the same in all inertial reference frames. In particular, if we are interested in galaxy correlations at some scale $k$, then large-scale perturbations at wave numbers $k'\ll k$ can only affect the statistics of the galaxy field via the local density or tidal field. The displacement (or velocity, or gravitational field -- these are all related in cosmological perturbation theory) can have no effect. Mathematically, since the displacement field in linear perturbation theory in Fourier space is ${\boldsymbol \Psi}(\mathbf k') = i(\mathbf k'/k'{^2})\delta(\mathbf k')$, effects of the displacement field are characterized by inverse powers of $k'$. Terms that have these inverse powers of $k'$ in the long-wavelength limit must cancel in any physical theory.

To see how Galilean invariance works in the case of streaming velocities, we consider the effect of large-scale power, only present up to some long wavelength $k_L$, on the observed tracer correlations at $k\gg k_L$. From Eq.~(\ref{eqn:pks}) and the non-symmetrized $F_2$, we see that $P_{\delta v^2}$ contains a term:
\begin{align}
P_{\delta v^2}(k) \supset &-2 \int  \frac{d ^3 \mathbf{k_1}}{(2 \pi)^3}   \mu_{12}^2 \frac{k_2}{k_1} T_v(k_1) T_v(k_2) \nonumber \\ & \times
\Pl(k_1) \Pl(k_2).
\end{align}
For $k_1\ll k$, $k_2 \approx k$, and the dependence on $\mu_{12}$ can be trivially averaged, yielding the contribution:
\linebreak
\begin{align}
P_{\delta v^2}(k) \supset &-\frac{2}{3} \int  \frac{d ^3 \mathbf{k_1}}{(2 \pi)^3} \frac{k}{k_1} T_v(k_1) T_v(k) \nonumber \\ & \times
\Pl(k_1) \Pl(k).
\label{eqn:divergence}
\end{align}
The $k_1^{-1}$ factor is problematic, as it suggests that the displacement field from the long-wavelength mode $k_1<k_L$ is affecting galaxy power at scale $k$. This implies that, in order to be physical, Eq.~(\ref{P_g}) must have another term proportional to $b_1b_v$ that cancels this divergence. The only other candidate is the advection term $P_{\rm adv}$; plugging Eq.~(\ref{eqn:L_s}) into Eq.~(\ref{eqn:pks}), we see that the advection term is
\begin{align}
P_{\rm adv}(k) \supset \frac{2}{3}\left[ \int  \frac{d ^3 \mathbf{k_1}}{(2 \pi)^3} \frac{T_v(k_1)}{k_1} 
\Pl(k_1)\right] kT_v(k) \Pl(k).
\end{align}
Thus the sum $P_{\delta v^2} + P_{\rm adv}$ contains no inverse powers of $k_1$ and obeys Galilean invariance, although each term does not individually. This cancellation is analogous to the one that occurs in the sum $P_{22}+P_{13}$ in the one-loop SPT power spectrum.

A subtlety in this argument is that at small $k_1$, $T_v(k_1)\propto k_1$, since the relative velocities of baryons and dark matter are sourced by the gradient in photon pressure in the early Universe (we have checked this scaling numerically from CLASS outputs). Thus the inverse power of $k_1$ in Eq.~(\ref{eqn:divergence}) is not realized in practice. However, the arguments in this appendix should be valid irrespective of the origin of the streaming velocities and hence the functional form of $T_v$.

\end{document}